\documentclass{article}
\newif\ifarxiv
\arxivtrue   

\usepackage{microtype}
\usepackage{graphicx}
\usepackage{subcaption}
\usepackage{booktabs}
\usepackage{hyperref}

\makeatletter
\def\input@path{{icml2026/}}
\makeatother

\ifarxiv
\usepackage[preprint]{icml2026}
\else
\usepackage{icml2026}
\fi

\usepackage{amsmath}
\usepackage{amsfonts}
\usepackage{amssymb}
\usepackage{comment}

\makeatletter
\expandafter\def\csname bibstyle@icml2026/icml2026\endcsname{\bibpunct{(}{)}{;}{a}{,}{,}}
\makeatother

\icmltitlerunning{Flash-SD-KDE}

\begin{document}

\twocolumn[
  \icmltitle{Flash-SD-KDE: Accelerating SD-KDE with Tensor Cores}

  \begin{icmlauthorlist}
    \icmlauthor{Elliot L.\ Epstein}{stanford}
    \icmlauthor{Rajat Vadiraj Dwaraknath}{stanford}
    \icmlauthor{John Winnicki}{stanford}
  \end{icmlauthorlist}

  \icmlaffiliation{stanford}{Institute for Computational and Mathematical Engineering, Stanford University}
  \icmlcorrespondingauthor{Elliot L.\ Epstein}{epsteine@stanford.edu}

  \icmlkeywords{kernel density estimation, GPU kernels, SD-KDE}
  \vskip 0.3in
]
\printAffiliationsAndNotice{}  

\begin{abstract}
Score-debiased kernel density estimation (SD-KDE) achieves improved asymptotic convergence rates over classical KDE, but its use of an empirical score has made it significantly slower in practice. 
We show that by re-ordering the SD-KDE computation to expose matrix-multiplication structure, Tensor Cores can be used to accelerate the GPU implementation. 
On a 32k-sample 16-dimensional problem, our approach runs up to $47\times$ faster than a strong SD-KDE GPU baseline and $3{,}300\times$ faster than scikit-learn's KDE. 
On a larger 1M-sample 16-dimensional task evaluated on 131k queries, Flash-SD-KDE completes in $2.3$ s on a single GPU, making score-debiased density estimation practical at previously infeasible scales.
\ificmlshowauthors Code available at \url{https://github.com/Elliotepsteino/Flash-SD-KDE}.\else Code available in supplementary material.\fi

\end{abstract}

\section{Introduction}

Score-debiased kernel density estimation (SD-KDE)~\citep{epstein2025sdkde}
improves the statistical efficiency of standard kernel density estimators,
relative to classical kernel density estimation (KDE)~\citep{rosenblatt1956remarks,parzen1962estimation,silverman1986density}:
for sufficiently smooth densities, SD-KDE attains better bias and mean-squared
error rates than vanilla KDE while retaining a simple, nonparametric form, while preserving nonnegativity.
In particular, SD-KDE achieves an asymptotic mean integrated squared error
(AMISE) of order $O(n^{-8/(d+8)})$, improving the convergence exponent
relative to the $O(n^{-4/(d+4)})$ rate of classical KDE tuned with
Silverman's rule of thumb~\citep{silverman1986density,wand1995kernel}.
These accuracy gains come with a significant computational drawback: the
empirical score used for debiasing introduces an additional $O(n^2)$ pass
over the data, so a naive implementation has roughly the same quadratic cost
as KDE itself but with a larger constant factor.

Given samples $x_i \in \mathbb{R}^d$ and a bandwidth $h$, a Gaussian KDE is
\[
  \hat p(x)
  =
  \frac{1}{n h^d}
  \sum_{i=1}^n
  \varphi\!\left(\frac{x - x_i}{h}\right),
\]
and SD-KDE forms debiased samples
$x_i^{\mathrm{SD}} = x_i + \tfrac{h^2}{2}\,\hat s(x_i)$ where
$\hat s$ is an estimate of the score.  In this work we always use an
empirical score computed from the KDE itself, for SD-KDE.  Assuming a standard Gaussian kernel, $\varphi$, writing the KDE explicitly in terms
of the samples, we estimate the score as
\[
  \hat s(x)
  =
  \frac{\nabla_x \hat p(x)}{\hat p(x)}
  =
  \frac{\sum_{i=1}^n \bigl[-({x - x_i})\,\varphi\!\left(\tfrac{x - x_i}{h}\right)\bigr]}
       {h^2 \sum_{i=1}^n \varphi\!\left(\tfrac{x - x_i}{h}\right)}.
\]

\section{Related Work}

\paragraph{Kernel density estimation.}
Kernel density estimation (KDE) is a classical nonparametric approach to density estimation that dates back to the foundational work of \citet{rosenblatt1956remarks} and \citet{parzen1962estimation}; modern treatments appear in standard references such as \citet{silverman1986density} and \citet{wand1995kernel}.
For smooth densities, KDE attains optimal nonparametric rates under appropriate bandwidth scaling, but in practice performance is highly sensitive to bandwidth selection.
Rules-of-thumb and plug-in procedures are widely used for their simplicity, but they are typically tuned to unimodal, near-Gaussian targets and can misbehave for multimodal or heavy-tailed distributions \citep{silverman1986density, botev2010diffusion}.
This sensitivity is particularly pronounced in moderate and high dimensions, where the bandwidth controls both statistical bias and the computational cost of evaluating kernels at scale.

\paragraph{Bias reduction and adaptive smoothing.}
A large body of work improves KDE accuracy by reducing leading-order bias or adapting the effective smoothing scale across the sample space.
One line of work constructs higher-order bias corrections by modifying kernels or combining estimators so that the dominant $O(h^2)$ bias term cancels, yielding $O(h^4)$ bias under additional smoothness assumptions \citep{fan1992bias, jones1997comparison}.
Another class of methods uses variable bandwidths that shrink in regions of high density and expand in the tails, improving adaptivity without committing to a parametric form \citep{abramson1982bandwidth, silverman1986density}.
Complementary to these ideas, diffusion-based estimators view density estimation through the lens of smoothing by a linear diffusion process, providing both adaptivity and a data-driven bandwidth selection mechanism \citep{botev2010diffusion}.
Score-debiased KDE (SD-KDE) \citep{epstein2025sdkde} fits into this landscape by using score information to correct the leading bias term while retaining a simple kernel estimator structure.
Our Laplace-corrected formulation can be seen as an explicit higher-order correction that removes the need to form an empirical score at evaluation time, trading additional kernel evaluations for improved numerical and computational behavior.

\paragraph{Score estimation and score-based modeling.}
The score function $\nabla_x \log p(x)$ is a central object in statistics and machine learning.
Score matching \citep{hyvarinen2005score} provides a way to fit unnormalized models by matching scores rather than likelihoods, and denoising score matching connects score estimation to learning denoisers \citep{vincent2011connection}.
More recently, score-based generative models learn scores of noise-perturbed data distributions with neural networks and sample via Langevin or SDE-based dynamics \citep{song2019generative, song2021sde, ho2020ddpm}.
These works typically target high-dimensional perceptual data and emphasize sampling or likelihood computation, whereas SD-KDE uses an explicit nonparametric density estimator and an analytically-defined score to improve pointwise density estimation.
Related to our setting, Stein-based score estimators recover score information from samples via identities that avoid explicit density evaluation \citep{li2017gradient}.
Our focus is complementary: we keep the statistical estimator fixed (SD-KDE and its Laplace variant) and address the practical bottleneck that arises when score-corrected estimators are deployed at large scale.

\paragraph{Fast evaluation of kernel sums and density models.}
The computational cost of KDE and related kernel methods is dominated by evaluating large Gaussian sums or kernel matrices.
Algorithmic accelerations include multipole and series-expansion methods such as the Fast Gauss Transform (FGT) \citep{greengard1991fast} and improved variants tailored to Gaussian kernels \citep{yang2003improved}.
A different set of approaches uses randomized low-rank approximations (e.g., random Fourier features) to trade bias for computational efficiency \citep{rahimi2007random}.
These methods reduce asymptotic complexity but can be sensitive to dimensionality, bandwidth, and target accuracy.
In contrast, our work targets the exact SD-KDE computation and focuses on constant-factor reductions that make the estimator practical on modern accelerators.

\paragraph{GPU acceleration and tensor-core programming.}
A growing ecosystem of libraries and DSLs targets large kernel computations on GPUs.
KeOps provides a memory-efficient reduction framework for kernel and distance matrices and is a strong baseline for Gaussian kernel reductions at scale \citep{charlier2021keops}.
General deep learning frameworks such as PyTorch \citep{paszke2019pytorch} make it easy to express quadratic kernel computations, but their performance depends critically on whether the computation can be lowered to high-throughput primitives.
Domain-specific languages such as Triton \citep{tillet2019triton} enable writing custom GPU kernels while still benefiting from compiler support for tiling and scheduling.
At the hardware level, modern GPUs provide specialized matrix-multiply units (Tensor Cores) and mixed-precision execution paths that can offer large throughput improvements when computations are expressed in GEMM-like form \citep{micikevicius2017mixed, williams2009roofline}.
Recent ``flash''-style algorithms demonstrate that reordering computations to avoid materializing large intermediate matrices and to match the GPU memory hierarchy can yield substantial speedups for quadratic primitives \citep{dao2022flashattention}.
Our contribution follows this direction for score-corrected density estimation: we reorder SD-KDE to expose matrix-multiplication structure and implement a streaming accumulation strategy so that Tensor Cores can be used effectively without incurring prohibitive memory traffic.

\section{Hardware}
\begin{figure*}[t]
  \centering
  \includegraphics[width=\textwidth]{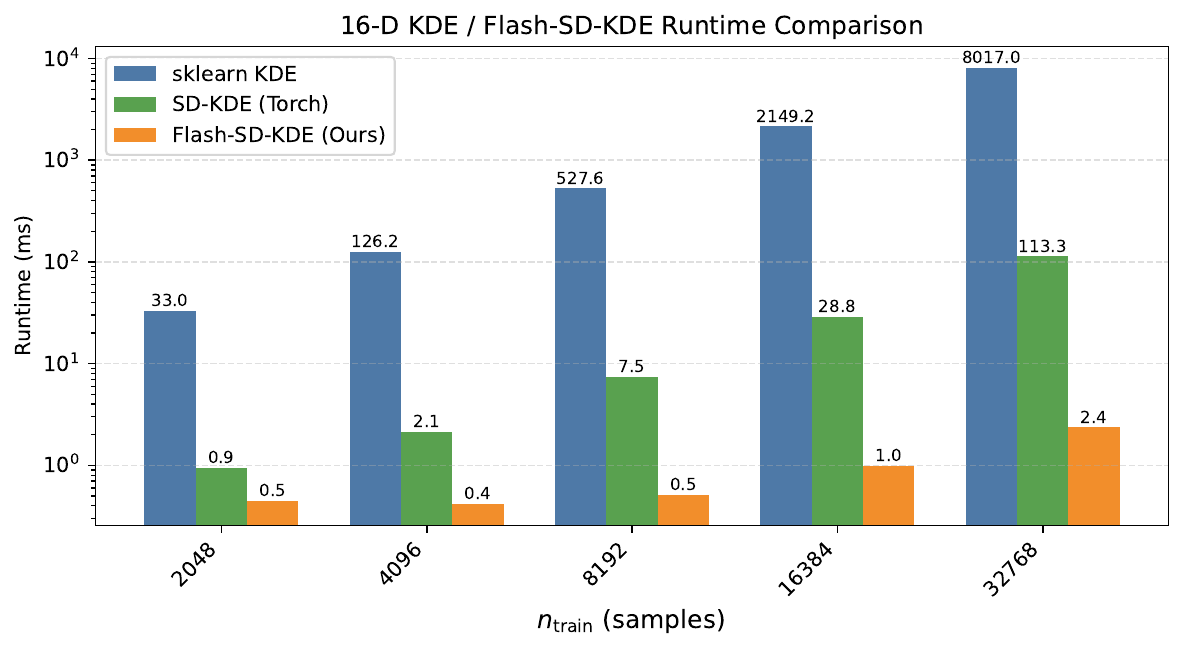}
  \caption{Runtime comparison for 16-D KDE/SD-KDE across $n_{\text{train}}$
           up to $32{,}768$ ($n_{\text{test}} = n_{\text{train}}/8$).}
  \label{fig:nd_runtime}
\end{figure*}
All experiments run on a workstation equipped with an NVIDIA RTX~A6000
(GA102).  This GPU exposes 84 streaming multiprocessors (SMs); each SM
contains 128 FP32 ALUs and 16 special function units (SFUs).  Because the
ratio of FP32 ALUs to SFUs is $128{:}16$, we treat one $\exp$ (issued on an
SFU) as costing the equivalent of $128/16 = 8$ FP32 flops in our models.
The card delivers roughly $40$~TFLOP/s of peak FP32 throughput and
approximately $770$~GB/s of GDDR6 bandwidth.

The host CPU is a dual-socket AMD EPYC~7763 system (``Milan'') configured
with $2 \times 64$ cores and two hardware threads per core (256 logical
CPUs reported by \texttt{lscpu}).  The processors boost up to
3.53~GHz, expose 512~MiB of shared L3 cache across the sockets, and provide
modern ISA extensions (AVX/AVX2, FMA, BMI1/2, SHA, VAES, etc.).

\section{Method}
\label{sec:method}

Our goal is to take $n_{\text{train}}$ samples in $d$ dimensions, form the
empirical SD-KDE (score + shift), and evaluate the resulting density on
$n_{\text{test}}$ queries.  Writing the computation in matrix form highlights
the parallel structure.  Let $x_i \in \mathbb{R}^d$ denote training points and $y_j \in \mathbb{R}^d$
denote query points, with bandwidth $h$ and squared Euclidean distance
\[
  \lVert x_i - y_j \rVert^2
  =
  \lVert x_i \rVert^2
  +
  \lVert y_j \rVert^2
  - 2\,x_i^\top y_j.
\]
Stacking the training samples into $X \in \mathbb{R}^{n_{\text{train}}\times d}$,
the empirical score computation is dominated by the train--train
Gram matrix
\[
  G_{\text{score}}
  =
  X X^\top
  \in
  \mathbb{R}^{n_{\text{train}}\times n_{\text{train}}}.
\]
Stacking queries into $Y \in \mathbb{R}^{n_{\text{test}}\times d}$ and
denoting the debiased training samples by
\[
  X^{\mathrm{SD}} = X + \tfrac{h^2}{2}\,\hat s(X),
\]
the final SD-KDE evaluation uses the train--query Gram matrix
\[
  G_{\text{KDE}}
  =
  X^{\mathrm{SD}} Y^\top
  \in
  \mathbb{R}^{n_{\text{train}}\times n_{\text{test}}}.
\]

On
modern NVIDIA GPUs this GEMM can be mapped to Tensor Cores~\citep{micikevicius2017mixed} and evaluated at
5--10$\times$ the throughput of standard FP32 SIMT arithmetic, particularly
when we restrict to $d$ that are multiples of 16 and use Triton's
\texttt{tl.dot} interface~\citep{tillet2019triton}.  The remaining operations---vector norms,
broadcasted additions, and exponentials---are all $O(n_{\text{train}}
 n_{\text{test}})$ but quickly become a small fraction of the total FLOPs as
$d$ grows.

The numerator in the empirical score inherits a similar GEMM structure.  Its numerator
involves terms of the form
\[
  \sum_j -(x_i - x_j)\,\varphi_{ij},
  \qquad
  \varphi_{ij}
  = \exp\!\left(-\frac{\lVert x_i - x_j\rVert^2}{2h^2}\right),
\]
which naively suggests $O(n_{\text{train}} n_{\text{test}} d)$ additional
elementwise arithmetic.  Using the identity
\[
  \sum_j (x_i - x_j)\,\varphi_{ij}
  =
  x_i \sum_j \varphi_{ij}
  -
  \sum_j \varphi_{ij}\,x_j,
\]
we can decompose the first term in the numerator into a elementwise sum, and the second term in the numerator into a second  operation:
\[
  T
  =
  \Phi X.
\]
Thus both the KDE evaluation and the SD-KDE score numerator reduce to
Tensor-Core-accelerable matrix multiplies, plus $O(n^2)$ scalar work for the
norms and exponentials.  In this paper we focus on the $d=16$ case, which
aligns naturally with the Tensor Core tile sizes on the RTX~A6000.

\subsection{Arithmetic intensity in $d$ dimensions}

To understand when the high-dimensional SD-KDE becomes compute-bound, we
estimate FLOPs, bytes moved, and arithmetic intensity for the $d$-dimensional
case.  Let $n_{\text{train}} = k$ and set $n_{\text{test}} = k/8$ as in the
experiments.

\paragraph{Total FLOPs.}
The d-dimensional implementation consists of three main matrix-multiply stages:
\begin{enumerate}
  \item Score Gram matrix $G_{\text{score}} = X X^\top$:
        $2 d k^2$ FLOPs.
  \item Score numerator $T = \Phi X$:
        $2 d k^2$ FLOPs, plus $4 k^2$ scalar FLOPs for norms and distance
        terms and $8 k^2$ FLOPs for exponentials (counting each $\exp$ as
        $8$ FLOPs due to the SFU/FP32 ratio).
  \item Final KDE Gram matrix on debiased data:
        $2 d k (k/8)$ FLOPs, plus $4 k (k/8)$ scalar FLOPs and
        $8 k (k/8)$ FLOPs for exponentials.
\end{enumerate}
Aggregating these terms yields
\[
  \begin{aligned}
    \text{FLOPs}_d(k)
    &\approx
    4 d k^2 + 12 k^2 + 2 d \frac{k^2}{8} + 12 \frac{k^2}{8} \\
    &=
    \Bigl(4 d + 12 + \tfrac{d}{4} + \tfrac{3}{2}\Bigr) k^2.
  \end{aligned}
\]
Specifically for the $d=16$ dimensional case, substituting $d=16$ gives
\[
  \text{FLOPs}_{16}(k)
  \approx
  81.5\,k^2,
\]
which is on the order of $10^{11}$ FLOPs for $k=32\text{k}$.

\paragraph{Bytes moved.}
To better capture implementation details, we count bytes per tile using the
launch parameters that delivered the best runtime ($BLOCK\_M = 64$,
$BLOCK\_N = 1024$).  Each tile loads $64\times d$ query values once
($4\,BLOCK\_M d$ bytes), streams $1024\times d$ training values
($4\,BLOCK\_N d$ bytes), and writes the partial PDF and weighted sums
($4\,BLOCK\_M$ and $4\,BLOCK\_M d$ bytes).  All of this traffic goes to and from
GDDR6, so for $d=16$ a tile moves roughly
\[
  \begin{aligned}
    \text{Bytes}_{\text{tile}}
    &\approx
    4\bigl(BLOCK\_M d + BLOCK\_N d \\
    &\qquad + BLOCK\_M + BLOCK\_M d\bigr) \\
    &= 4\bigl(2\,BLOCK\_M d + BLOCK\_N d \\
    &\qquad + BLOCK\_M\bigr) \\
    &\approx 7.4\times 10^4\ \text{bytes}.
  \end{aligned}
\]
The full problem processes $(k / BLOCK\_M)\times(k / BLOCK\_N)$ such tiles,
so the total GDDR6 traffic is
\[
  \begin{aligned}
    \text{Bytes}_{16}(k)
    &\approx
    \text{Bytes}_{\text{tile}}
    \times
    \frac{k}{BLOCK\_M}
    \times
    \frac{k}{BLOCK\_N} \\
    &\approx
    1.13\,k^2\ \text{bytes}.
  \end{aligned}
\]

\paragraph{Arithmetic intensity.}
Dividing FLOPs by bytes gives
\[
  \begin{aligned}
    I_d(k)
    &=
    \frac{\text{FLOPs}_d(k)}{\text{Bytes}_d(k)} \\
    &\approx
    \frac{\bigl(4 d + 12 + \tfrac{d}{4} + \tfrac{3}{2}\bigr) k^2}
         {4\bigl(\tfrac{9}{8} d k + \tfrac{k}{8}\bigr)} \\
    &\sim
    C(d)\,k
    \quad\text{for large }k,
  \end{aligned}
\]
with
\[
  C(d)
  \approx
  \frac{4 d + 12 + d/4 + 3/2}{4\cdot (9 d/8)}
  =
  \frac{(17/4)\,d + 27/2}{9 d/2}.
\]
For $d=16$ this simplifies to
\[
  I_{16}(k)
  \approx
  \frac{\text{FLOPs}_{16}(k)}{\text{Bytes}_{16}(k)}
  \approx
  \frac{81.5\,k^2}{1.13\,k^2}
  \approx 72\ \text{flops/byte}.
\]
This tile-aware estimate more closely reflects the measured arithmetic
intensity.  Comparing against the A6000 specs (Tensor Core peak of
$155\,\text{TFLOP/s}$ versus $\approx 770\,\text{GB/s}$ of memory bandwidth),
the machine balance is roughly $200$ flops/byte~\citep{williams2009roofline}.  Since our kernel sustains
over $70$ flops/byte, it lies well into the compute-bound regime on the
RTX~A6000.  Nsight Compute reports an empirical arithmetic intensity of roughly
$95\,\text{FLOPs/byte}$ for the score kernel on the A6000, which is broadly in
line with the $72\,\text{FLOPs/byte}$ predicted by the tile-level model above.
Minor differences stem from additional bookkeeping work (atomics, reductions)
and from Nsight's finer-grained accounting of texture/cache traffic, but both
figures agree that the kernel operates far above the bandwidth roofline.
Because the kernel mixes Tensor-Core GEMMs with FP32 scalar work (norms,
exponentials, atomics), the effective machine balance sits between the
tensor-core roof ($\approx 200$ flops/byte) and the FP32 roof ($\approx 50$
flops/byte).  The observed $70$--$95$ flops/byte therefore straddle these two
limits: the GEMM portion is partially bandwidth-limited relative to Tensor
Cores, while the scalar portion is compute-limited relative to FP32 ALUs.

\section{Laplace-corrected KDE}
\label{sec:laplace_corrected}
\begin{figure*}[t]
  \centering
  \includegraphics[width=\textwidth]{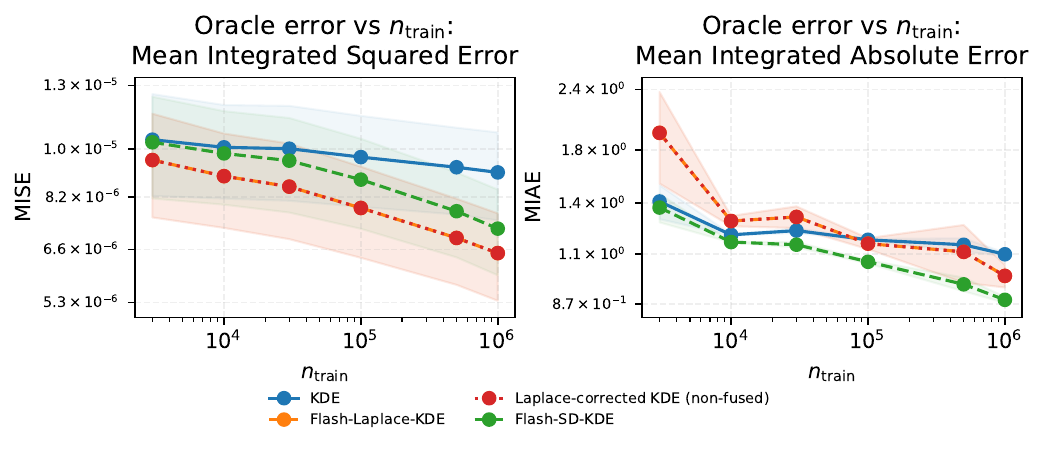}
  \caption{Oracle error on a 16D mixture-of-Gaussians benchmark.
           We report MISE and MIAE versus $n_{\text{train}}$ for KDE, Flash-Laplace-KDE (fused Laplace correction), non-fused Laplace correction, and Flash-SD-KDE.
           The Laplace-corrected estimators can be slightly negative, so error is computed in a signed density manner.}
  \label{fig:oracle_16d_error_1d}
\end{figure*}

\begin{figure*}[t]
  \centering
  \includegraphics[width=\textwidth]{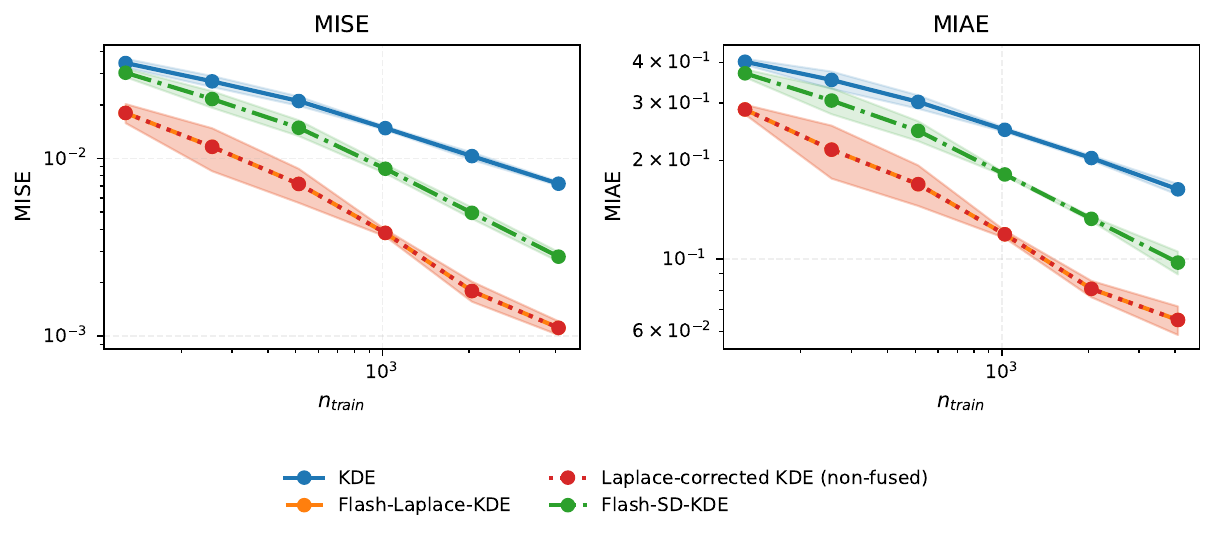}
  \caption{Oracle error on a 1D mixture-of-Gaussians benchmark.
           We report MISE and MIAE versus $n_{\text{train}}$ for KDE, Flash-Laplace-KDE (fused Laplace correction), non-fused Laplace correction, and Flash-SD-KDE.
           The Laplace-corrected estimators can be slightly negative, so error is computed on the signed density.}
  \label{fig:oracle_error_1d}
\end{figure*}

\begin{figure*}[t]
  \centering
  \includegraphics[width=\textwidth]{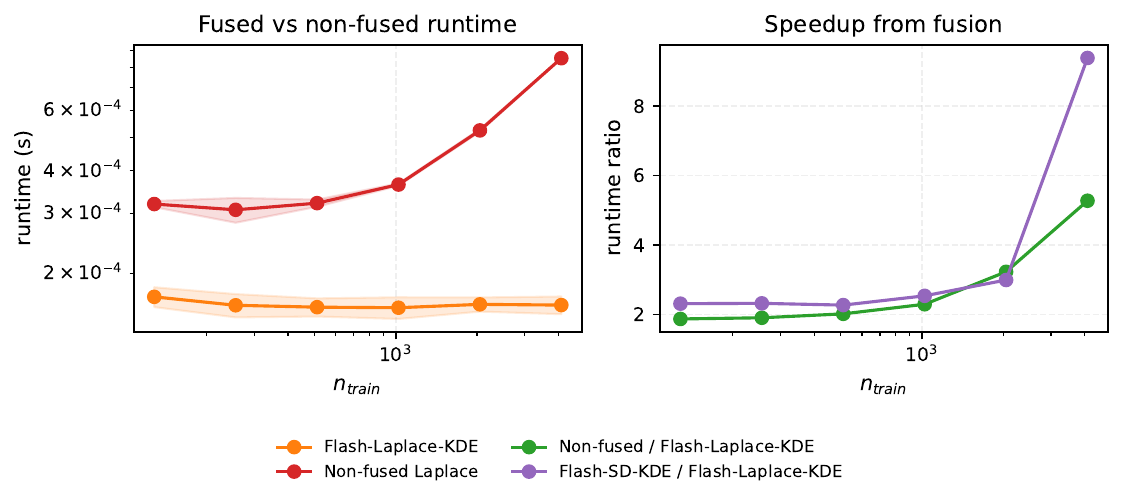}
  \caption{Runtime and speedup for Laplace correction in 1D.
           The left panel shows total runtime for the fused Flash-Laplace-KDE kernel and a non-fused implementation.
           The right panel reports speedup ratios, including Flash-SD-KDE relative to Flash-Laplace-KDE for context.}
  \label{fig:laplace_speedup_1d}
\end{figure*}
Score-debiased KDE offers better asymptotic error rates, but its empirical score step adds a full quadratic pass over the data.
A natural approximation is to replace the explicit debiasing with a correction to the kernel itself, in the spirit of classical higher-order bias corrections for KDE\footnote{Laplace kernel has $0$ second moment, but non-zero fourth moment, so it is a fourth-order kernel.}~\citep{fan1992bias,jones1997comparison}.
This yields a Laplace-corrected KDE that captures the same leading-order bias reduction of SD-KDE~\citep{epstein2025sdkde} while avoiding the empirical score computation at the cost of admitting possibly negative values.

Let $K_h(x)$ denote the isotropic Gaussian kernel in $d$ dimensions.
\[
  K_h(x)
  =
  \frac{1}{(2\pi)^{d/2} h^d}
  \exp\!\left(-\frac{\lVert x \rVert^2}{2h^2}\right),
\]
Let $\hat p(x) = \tfrac{1}{n} \sum_i K_h(x - x_i)$ be the corresponding KDE.
The Laplace-corrected kernel is defined by
\[
  K_h^{\text{LC}}(x)
  =
  K_h(x) - \frac{h^2}{2} \Delta K_h(x).
\]
For the Gaussian kernel, the Laplacian has a closed form.
\[
  K_h^{\text{LC}}(x)
  =
  K_h(x)\left(1 + \frac{d}{2} - \frac{\lVert x \rVert^2}{2h^2}\right),
\]
The estimator becomes
\[
  \hat p_{\text{LC}}(x)
  =
  \frac{1}{n} \sum_{i=1}^n K_h^{\text{LC}}(x - x_i).
\]

\paragraph{Connection to SD-KDE.}
SD-KDE~\citep{epstein2025sdkde} forms debiased samples $x_i^{\text{SD}} = x_i + \tfrac{h^2}{2}\,\hat s(x_i)$ using an empirical score $\hat s$.
Evaluating a KDE on the shifted samples and linearizing in $h^2$ yields a correction proportional to $\Delta K_h$.
When $\hat s$ is taken to be the KDE score, this linearized estimator coincides with the Laplace-corrected kernel above.
Thus $\hat p_{\text{LC}}$ can be viewed as a first-order approximation to SD-KDE that preserves its leading bias reduction while avoiding an explicit score pass.

As an intuitive explanation of why both methods achieve the same leading bias reduction, we can first view vanilla KDE\footnote{Throughout, we assume that the kernel is Gaussian for simplicity.} as a heat semigroup (kernel) $P_{t} = e^{\frac{t}{2}\Delta}$ with $t = h^2$ applying to the empirical density  function $f_n$ (sampled with replacement) to be
\begin{align*}
    \hat{f} = P_t f_n.
\end{align*}
From the linearity of the heat semigroup, we then have that the expected estimate
\begin{align*}
    \mathbb{E}\left[\hat{f}\right]
    =\mathbb{E}\left[P_t f_n\right]
    =P_t\mathbb{E}\left[ f_n\right]
    =P_tf,
\end{align*}
where $f$ is the true probability density function.

The bias is then, under a smoothness assumption we assume throughout,
\begin{align*}
    \mathbb{E}\left[\hat{f}\right]-f = (P_t-I)f = \frac{t}{2} \Delta f + O(t^2).
\end{align*}

The Laplace kernel deals with this leading term by using an operator $\left(I-\frac{t}{2}\Delta\right)P_t$ instead, making its bias
\begin{align*}
    \mathbb{E}\left[\hat{f}^{\text{LC}}\right]-f = \left[\left(I-\frac{t}{2}\Delta\right)P_t-I\right]f = O(t^2).
\end{align*}

The SD-KDE algorithm approaches this debias task differently by applying a \textbf{non-linear}\footnote{Even when the transportation kernel has a non-linear transportation. The kernel itself is linear.} transportation kernel $T_{\frac{t}{2}\nabla \log f}$ before applying a standard heat kernel $P_t$, making its bias
\begin{align*}
    &\mathbb{E}\left[\hat{f}^{\text{SD-KDE}}\right]-f\\ &= \left[P_tT_{\frac{t}{2}\nabla \log f}-I\right]f\\
    &=
    \left[P_t\left(f - \frac{t}{2}\nabla \cdot \left(f\nabla \log f\right) + O(t^2)\right)-f\right]\\
    &=
    \left[P_t\left(f - \frac{t}{2}\Delta f + O(t^2)\right)-f\right]\\
    &=
    \left[P_t\left(I - \frac{t}{2}\Delta  \right)-I\right]f+ O(t^2)\\
    &= O(t^2).
\end{align*}
Recall that $t=h^2$, so we have that Laplace KDE and SD-KDE have an $O(h^4)$ bias, while vanilla KDE has an $O(h^2)$ bias.

Note that the analysis above assumes that the SD-KDE has access to the oracle, so the translation kernel is linear. In practice, the score is empirical, making the translation kernel
\begin{align*}
    T_{\frac{t}{2}\nabla \log \left(P_{t'} f_n\right)}
\end{align*}
where $t'=\frac{h^2}{2}$ is the bandwidth for the kernel used for score estimation. Thus, the empirical SD-KDE is
\begin{align*}
    \hat{f}^{\text{Emp SD-KDE}} = T_{\frac{t}{2}\nabla \log \left(P_{t'} f_n\right)} f_n,
\end{align*}
which is no longer linear in $f_n$.

\paragraph{Why we consider this correction.}
The Laplace-corrected estimator is appealing when the full SD-KDE score is too expensive or when a fast bias-reduction surrogate is sufficient.
It requires only the same pairwise distances and exponentials as standard KDE, plus a simple affine factor in the kernel value.
The correction can be negative for large $\lVert x - x_i \rVert$, so $\hat p_{\text{LC}}$ is not guaranteed to be nonnegative, which we account for in our experiments.

\paragraph{Kernel fusion opportunity.}
Computing $K_h^{\text{LC}}$ reuses the same scaled distances used to compute $K_h$.
A fused kernel can compute $\phi = \exp(-\tfrac{1}{2}\,\text{scaled})$ and apply the Laplace factor inside the same pass.
A non-fused implementation must either recompute distances or materialize intermediates in a second kernel.
This increases memory traffic and kernel launches.
We therefore treat the fused Laplace-corrected kernel as a distinct fast path and refer to it as \emph{Flash-Laplace-KDE}.

\section{Results}

We now summarize the empirical behavior of the 16-D implementation on the
RTX~A6000.  For each $n_{\text{train}}$ in $\{2\text{k}, 4\text{k}, 8\text{k}, 16\text{k},
32\text{k}\}$ we set $n_{\text{test}} = n_{\text{train}}/8$, draw data from a
simple 16-D Gaussian mixture, and run three baselines: scikit-learn KDE~\citep{pedregosa2011scikit},
Torch SD-KDE (GEMM-based, implemented in PyTorch~\citep{paszke2019pytorch}), and
Flash-SD-KDE (Tensor-Core GEMMs for both KDE and score, implemented in
Triton~\citep{tillet2019triton}).  Figure~\ref{fig:nd_runtime} shows that the Flash-SD-KDE rapidly
outpaces both baselines as $n$ grows.

We also measure utilization by combining the flop model above with the
measured runtimes.  As shown in Figure~\ref{fig:triton_sd_kde_nd_util}, the Flash-SD-KDE GPU utilization is high into the multi-digit range once
$n_{\text{train}}$ exceeds $8\text{k}$, despite a lot of operations such as exponential computations that can't utilize tensor cores.  This confirms that the 16-D
Tensor-Core formulation is firmly compute-bound and that additional tuning
effort should focus on kernel fusion and occupancy rather than memory traffic.

To compare against a specialized KDE implementation, we also benchmarked PyKeOps
LazyTensor operators at $n_{\text{train}}=32\text{k}$, $n_{\text{test}}=4\text{k}$.
PyKeOps currently represents the state of the art for large-scale kernel
reductions, so we report both its Gaussian KDE runtime and its SD-KDE runtime as
a reference point.  Table~\ref{tab:pykeops} shows that Flash-SD-KDE runs
$1.57\times$ faster than the PyKeOps 16-D KDE baseline and $7.99\times$ faster
than the PyKeOps 16-D SD-KDE implementation.

\begin{table}[h]
  \centering
  \small
  \setlength{\tabcolsep}{4pt}
  \begin{tabular}{lcc}
    \toprule
    Method & Runtime (ms) & Rel. to Flash-SD-KDE \\
    \midrule
    16-D Flash-SD-KDE & $2.11$ & $1\times$ \\
    PyKeOps 16-D KDE   & $3.33$ & $1.57\times$ \\
    PyKeOps 16-D SD-KDE& $16.91$ & $7.99\times$ \\
    \bottomrule
  \end{tabular}
  \caption{Runtime comparison at $n_{\text{train}}=32\text{k}$,
           $n_{\text{test}}=4\text{k}$.  PyKeOps rows report Gaussian KDE and
           SD-KDE (LazyTensor) runtimes, while Flash-SD-KDE executes the full SD-KDE
           pipeline.}
  \label{tab:pykeops}
\end{table}

\subsection{Oracle density benchmark for Laplace correction}
Figure~\ref{fig:oracle_16d_error_1d} reports oracle accuracy on the 16-D mixture benchmark as a function of $n_{\text{train}}$, using Mean Integrated Squared Error (MISE) on the left and Mean Integrated Absolute Error (MIAE) on the right. 
Across the sweep, Flash-SD-KDE substantially improves on vanilla KDE, indicating that the baseline KDE estimator is a poor fit in this regime. 
The Laplace-corrected variants achieve the lowest MISE throughout, and the fused Flash-Laplace-KDE curve overlaps the non-fused implementation, showing that fusion is an implementation optimization rather than a change in the estimator. 
In MIAE, Flash-SD-KDE attains the lowest error, while the Laplace-corrected estimators show a sharp drop between the smallest $n_{\text{train}}$ and $10^4$, followed by a small non-monotonic “kink” (a slight increase at intermediate $n_{\text{train}}$). 
The kink lies within the uncertainty bands and is consistent with finite-sample variability and the greater sensitivity of an absolute-error metric to the signed tail correction. 
Both metrics continue to decrease as $n_{\text{train}}$ grows. Because the Laplace-corrected kernel is not guaranteed to remain nonnegative, we compute MISE/MIAE on the signed estimator and separately log the integrated negative mass as a diagnostic.

We also evaluate Laplace-corrected KDE on a 1D oracle density.
Figure~\ref{fig:oracle_error_1d} compares KDE, Flash-Laplace-KDE (fused Laplace correction), non-fused Laplace correction, and Flash-SD-KDE.
The Laplace-corrected estimators can take small negative values for large $\lVert x - x_i \rVert$.
We therefore compute errors on the signed density, and we report the negative-mass diagnostics separately in the benchmark logs.

Figure~\ref{fig:laplace_speedup_1d} shows the runtime advantage of fusion.
The fused Flash-Laplace-KDE kernel consistently outperforms the non-fused Laplace correction across $n_{\text{train}}$.
For context, the same plot also reports the Flash-SD-KDE to Flash-Laplace-KDE runtime ratio.

\subsection{Performance optimizations}
\begin{figure*}[t]
  \centering
  \includegraphics[width=0.9\textwidth]{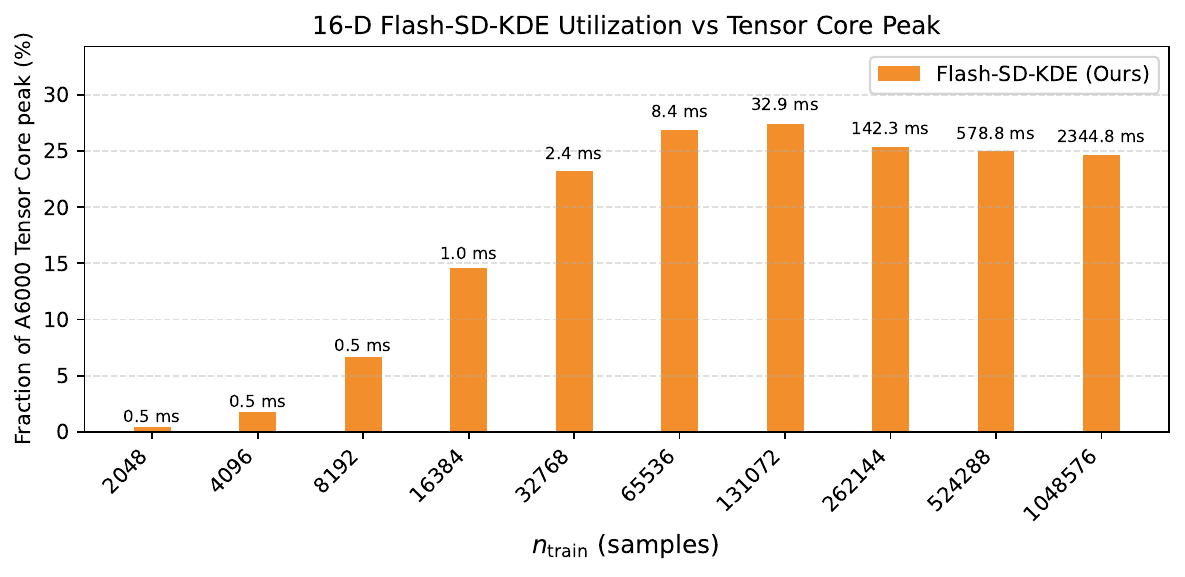}
  \caption{Utilization (percentage of RTX A6000 Tensor Core peak) for the 16-D
           SD-KDE pipeline, computed via the flop model from
           Section~\ref{sec:method}; bars are annotated with the observed
           runtimes (ms).}
  \label{fig:triton_sd_kde_nd_util}
\end{figure*}
Nsight Systems traces show that roughly $95\%$ of the SD-KDE runtime at
$n_{\text{train}}=32\text{k}$, $n_{\text{test}}=4\text{k}$ is spent computing
the empirical score.  We therefore concentrated optimization effort on the
score kernel, sweeping the launch parameters to raise utilization.  Specifically
we varied:
\begin{itemize}
  \item $BLOCK\_M \in \{32, 64, 128, 256\}$,
  \item $BLOCK\_N \in \{32, 64, 128, 256, 512, 1024\}$,
  \item $num\_warps \in \{1, 2, 4, 8\}$,
  \item $num\_stages \in \{1, 2, 4\}$,
\end{itemize}
and selected the combination that minimized runtime.  For this workload we
found that $BLOCK\_M=64$, $BLOCK\_N=1024$, $num\_warps=2$, and $num\_stages=2$
gave the best overall performance, increasing measured FLOP utilization by
more than $2\times$ relative to smaller-tile settings.

Beyond the launch-parameter sweep, two design choices account for most of the
speedup over a standard tiling implementation:
\begin{enumerate}
  \item \textbf{Tensor-Core tiling:} all high-dimensional dot products are
        processed as $16 \times 16$ tiles with Tensor Core
        acceleration, allowing the GEMM portion of the score and KDE kernels
        to run near the TF32 roofline.
  \item \textbf{Streaming accumulation:} we never materialize full
        $n_{\text{train}}\times n_{\text{train}}$ (or query) matrices; instead
        we stream tiles through registers and rely on atomic reductions to keep
        global-memory traffic linear in $n$.
\end{enumerate}
Together these optimizations push the kernels close to the hardware limit.
Nsight Compute's ``Speed of Light'' report shows $\sim 68\%$ SM throughput and
roughly $90\%$ L1 throughput for the score kernel at
$n_{\text{train}}=32\text{k}$ and $n_{\text{test}}=4\text{k}$, which is
consistent with the tensor-core utilization trends in
Figure~\ref{fig:triton_sd_kde_nd_util}: we cannot reach the nominal Tensor
Core peak because the kernel still executes substantial non-Tensor-Core work
(norms, exponentials, atomics).  At these utilization levels we stopped
further low-level tuning, since additional optimizations would likely yield
only modest incremental gains.

\label{sec:conclusion}
\section{Conclusion}
Score-debiased kernel density estimation (SD-KDE) offers improved asymptotic accuracy over classical KDE, but its reliance on an empirical score has historically made it substantially slower in practice. 
This paper shows that the dominant cost of empirical SD-KDE is not inherently prohibitive: by re-ordering the computation to expose matrix-multiplication structure, both KDE evaluation and the score numerator can be expressed as Tensor Core--accelerable GEMMs plus lightweight elementwise operations, while avoiding materializing full pairwise interaction matrices via streaming accumulation. 
In the 16-D setting, this hardware-aligned formulation yields large end-to-end gains---up to $47\times$ faster than a strong Torch SD-KDE baseline and $3{,}300\times$ faster than scikit-learn KDE---and scales to previously infeasible regimes, completing SD-KDE on $\sim$1M training points and $\sim$131k queries in 2.3\,s on a single GPU.

 Overall, our results suggest that practical nonparametric estimators can often be unlocked by hardware-aware reformulations that maximize GPU utilization. 
 Future directions include extending the approach beyond $d$ multiples of 16, supporting richer bandwidth parameterizations and kernels, further reducing non-Tensor-Core overheads (e.g., exponentials and atomics), and developing nonnegativity-preserving approximations and multi-GPU variants.

\ificmlshowauthors
\section{Acknowledgements}
The authors would like to thank Thanawat Sornwanee for helpful discussions.

\fi
\section*{Impact Statement}

This paper presents work whose goal is to accelerate SD-KDE and KDE kernels on
GPUs. We expect the primary impact to be enabling practitioners to apply these
estimators to larger datasets; we do not foresee unique societal consequences
outside the usual considerations for density estimation applications.

\bibliographystyle{icml2026/icml2026}
\bibliography{references}

\appendix
\section{Flash-SD-KDE in low dimensions}

For completeness we briefly summarize the 1-D SD-KDE arithmetic intensity and
empirical behavior.  With $n_{\text{train}} = k$ and $n_{\text{test}} = k/8$,
there are two steps:
\begin{enumerate}
  \item \textbf{Score + shift:} For each training point we compute
        $\hat s(x_i)$ from all $k$ points and then form
        $x_i^{\mathrm{SD}} = x_i + \tfrac{h^2}{2}\hat s(x_i)$.  This uses
        $O(k^2)$ pairwise kernel interactions.  With the RTX~A6000 hardware
        ratio ($128$ FP32 ALUs, $16$ SFUs per SM) we budget an $\exp$ as
        $8$ flop-equivalents.  The score accumulation requires one $\exp$
        and roughly eight additional arithmetic ops (subtraction, scaling,
        accumulation), yielding $c_1 \approx 16$ flops per
        $(\text{train},\text{train})$ pair.
  \item \textbf{KDE on debiased samples:} We then evaluate a standard Gaussian
        KDE at $k/8$ query points using the $k$ debiased samples, which uses
        $O(k^2/8)$ interactions.  Each pair requires one $\exp$ (8 flops) plus
        about six other operations (difference, square, scaling, accumulation),
        so we take $c_2 \approx 14$ flops per $(\text{train},\text{test})$ pair.
\end{enumerate}
The total work is therefore approximated by
\[
  \text{FLOPs}(k)
  \;\approx\;
  c_1 k^2 + c_2\,k\,(k/8)
  \;\approx\;
  16 k^2 + 14\frac{k^2}{8}
  \;=\;
  17.75\,k^2.
\]
For $k = 32\text{k}$ this is on the order of $2\times 10^{10}$ flops.  When we
fix $n_{\text{test}} = k/8$ we move approximately $5k$ bytes (counting one
read of each train and test point and one write of each output), so the
arithmetic intensity scales as
\[
  I(k)
  =
  \frac{\text{FLOPs}(k)}{\text{Bytes}(k)}
  \;\approx\;
  \frac{17.75\,k^2}{5\,k}
  \approx
  3.55\,k
  \quad \text{flops/byte},
\]
placing realistic problem sizes firmly in the compute-bound regime.

\begin{figure*}[th]
  \centering
  \includegraphics[width=0.8\linewidth]{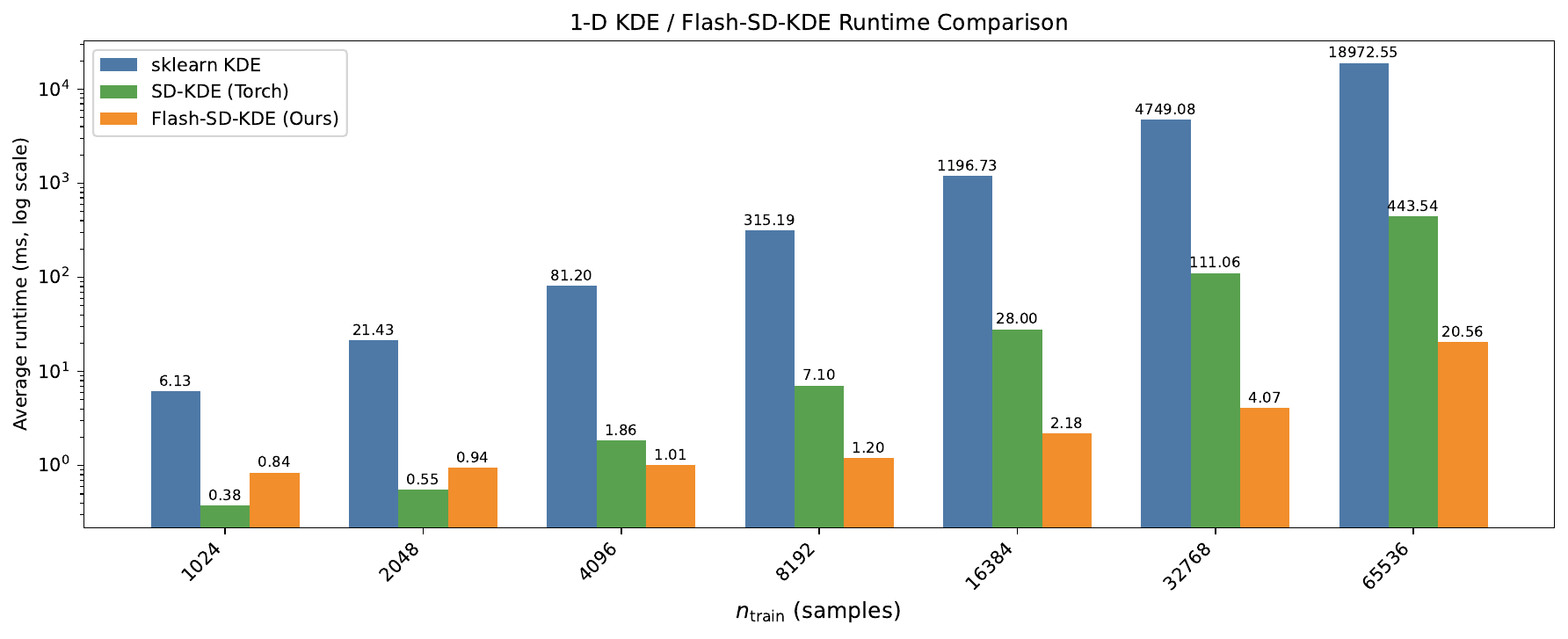}
  \caption{Average runtime (log-scale $y$-axis) of scikit-learn KDE, SD-KDE (Torch), and Flash-SD-KDE across $n_{\text{train}}$ in 1-D; annotations show observed runtimes.}
  \label{fig:flash_sd_kde}
\end{figure*}

Empirically we sweep $k$ over powers of two from $1024$ to $64\text{k}$, with
$n_{\text{test}} = k/8$, and average over three seeds.  For each
configuration we record scikit-learn Gaussian KDE
time, and SD-KDE Torch time, and Flash-SD-KDE time.  Across the entire range, the Flash-SD-KDE
implementation is consistently faster than scikit-learn, with speedups
growing with $k$; at the largest sizes, Flash-SD-KDE achieves around 2 orders-of-magnitude speedup relative to the sklearn baseline, as shown in Figure~\ref{fig:flash_sd_kde}.

\begin{figure*}[th]
  \centering
  \includegraphics[width=0.8\linewidth]{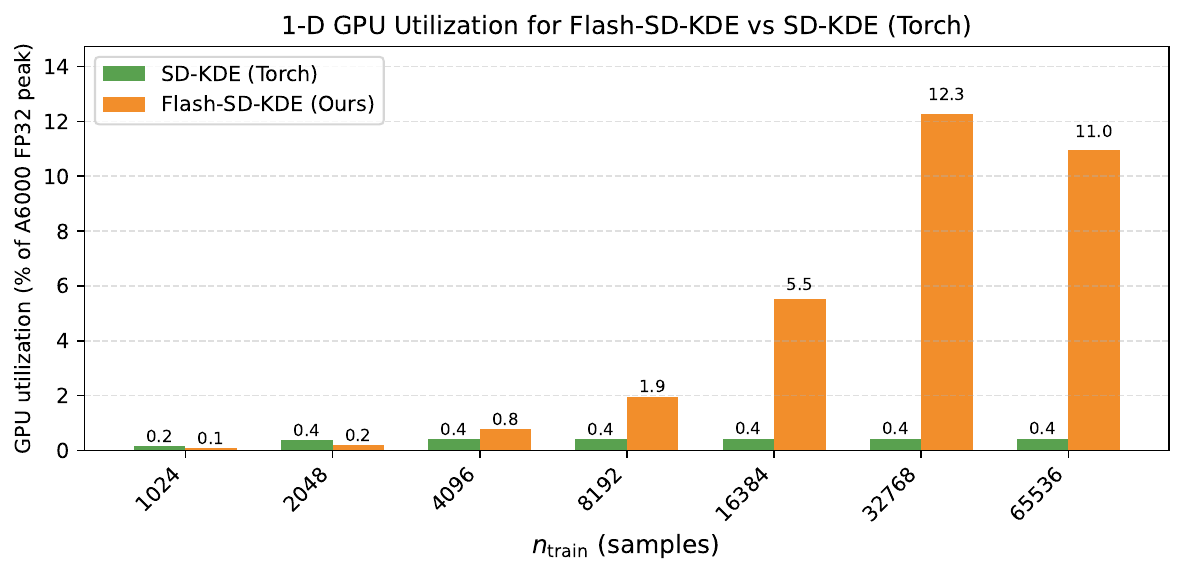}
  \caption{Utilization of the 1-D Flash-SD-KDE kernel and SD-KDE (Torch) for large
           $n_{\text{train}}$ (powers of two up to $2^{16}$).  Labels show
           the observed runtime for each data size; error bars are omitted
           because a single seed is used per point.}
  \label{fig:triton_large_util}
\end{figure*}

As shown in Figure~\ref{fig:triton_large_util}, we also compare the utilization of Flash-SD-KDE with SD-KDE (Torch) over a range of 1-D problem sizes, using powers of
two for $n_{\text{train}}$ up to $2^{16} \approx 64$ thousand (and
$n_{\text{test}} = n_{\text{train}}/8$).  

\end{document}